\documentclass[conference,twocolumn,10pt]{IEEEtran} 
\usepackage[utf8]{inputenc}
\usepackage{epsfig,cite}
\usepackage{graphicx}
\graphicspath{{graphics/}}
\usepackage{color}
\usepackage{amssymb,amsmath}
\usepackage{mathrsfs} 
\allowdisplaybreaks 

\newif\ifShowFullContent

\IEEEoverridecommandlockouts

\DeclareMathOperator{\erf}{erf}

\begin{document}
\bibliographystyle{IEEEtran}

\title{Root Mean Square Error of Neural Spike Train\\ Sequence Matching with Optogenetics}

\author{
	\IEEEauthorblockN{Adam Noel, Dimitrios Makrakis}
	\IEEEauthorblockA{School of EECS\\ University of Ottawa, Ottawa, Ontario, Canada\\ Emails: anoel2@uottawa.ca, dimitris@eecs.uottawa.ca}
	\and
	\IEEEauthorblockN{Andrew W. Eckford}
	\IEEEauthorblockA{Dept. of EECS\\
	York University,
	Toronto, Ontario, Canada\\
	Email: aeckford@yorku.ca}

\thanks{This work was supported in part by the Natural Sciences and Engineering Research Council of Canada (NSERC).}

}

\newcommand{\EXP}[1]{\exp\left(#1\right)}
\newcommand{\ERF}[1]{\erf\left(#1\right)}

\newcommand{\prob}[1]{P_\textnormal{#1}}

\newcommand{\threshold}{\tau}

\newcommand{\dt}{\Delta t}

\makeatletter
\newcommand{\vast}{\bBigg@{3}}
\newcommand{\Vast}{\bBigg@{4}}
\makeatother

\newcommand{\light}{\ell}
\newcommand{\rateTarget}{\lambda_\mathrm{T}}
\newcommand{\probTarget}{p_\mathrm{T}}
\newcommand{\fireThresh}{v_\textrm{fire}}
\newcommand{\tMin}{t_\textrm{min}}
\newcommand{\nMin}{{n_\textrm{min}}}
\newcommand{\tSep}{t_\Delta}
\newcommand{\nSep}{n_\Delta}
\newcommand{\distortion}{d}

\newcommand{\target}[1]{u_{#1}}
\newcommand{\targetSeq}{\vec{u}}
\newcommand{\targetLen}{M}
\newcommand{\gen}[1]{v_{#1}}
\newcommand{\genSeq}{\vec{v}}
\newcommand{\genCand}[1]{w_{#1}}
\newcommand{\genSeqCand}{\vec{w}}
\newcommand{\genLen}{N}
\newcommand{\offset}[1]{a_{#1}}
\newcommand{\offsetSeq}{\vec{a}}
\newcommand{\map}[1]{f\left(#1\right)}
\newcommand{\mapDT}[1]{f\left[#1\right]}
\newcommand{\kernel}[1]{h\left(#1\right)}
\newcommand{\kernelDT}[1]{h\left[#1\right]}
\newcommand{\kernelVal}[1]{h_{#1}}
\newcommand{\kernelDTSq}[1]{h^2\left[#1\right]}
\newcommand{\distortFcn}[1]{d\left(#1\right)}
\newcommand{\distortFcnDT}[1]{d\left(#1\right)}
\newcommand{\distortFcnDTAvg}[1]{\overline{d}\left(#1\right)}
\newcommand{\pulse}{\delta t}
\newcommand{\pulseDT}{\Delta W}

\newcommand{\bDelayRV}[1]{X_{#1}}
\newcommand{\bDelayRVval}[1]{x_{#1}}
\newcommand{\distortRV}{Y}
\newcommand{\distortRVval}{y}
\newcommand{\arbitraryRV}{Z}
\newcommand{\arbitraryRVval}{z}
\newcommand{\pmfFcn}[1]{p\left(#1\right)}

\newcommand{\ion}{I_{\mathrm{on}}}

\maketitle

\begin{abstract}
Optogenetics is an emerging field of neuroscience where neurons are genetically modified to express light-sensitive receptors that enable external control over when the neurons fire. Given the prominence of neuronal signaling within the brain and throughout the body, optogenetics has significant potential to improve the understanding of the nervous system and to develop treatments for neurological diseases. This paper uses a simple optogenetic model to compare the timing distortion between a randomly-generated target spike sequence and an externally-stimulated neuron spike sequence. The distortion is measured by filtering each sequence and finding the root mean square error between the two filter outputs. The expected distortion is derived in closed form when the target sequence generation rate is sufficiently low. Derivations are verified via simulations.
\end{abstract}

\maketitle

\section{Introduction}

The nervous system is the most complex system of the human body and understanding this system is considered to be one of the biggest challenges in all of biology; see \cite[Ch.~45]{Sadava2014}. The neural network, with up to $10^{14}$ connections within the brain, also controls bodily functions such as muscle contraction. The transfer of information is not entirely internal; sensory neurons, such as those in the retina, generate and propagate signals in response to external stimuli.

There is significant interest in developing methods to control the external excitation of neurons to improve our understanding of the nervous system and develop treatments for neurological diseases. One prominent example is the emerging field of optogenetics; see \cite{Fenno2011}. Optogenetics uses a relatively simple genetic modification to induce a neuron to express light-sensitive receptors on its membrane. These light-gated receptors can then be used to adjust the ion current across the membrane, which enables one to alter its electrical potential and control when it fires. Experiments in \cite{Nagel2002,Nagel2003} identified opsin-based receptors such as Channelrhodopsin (ChR) to be particularly suitable for optogenetic studies, due to its simplicity and its compatibility for implanting in \emph{living} animals, as first demonstrated in the worm \emph{C. elegans} in \cite{Nagel2005}.

From a communication perspective, nanoscale stimulators were proposed in \cite{Balasubramaniam2011,Mesiti2013} to control neurons and interface with a neural network. In \cite{Wirdatmadja2016}, it was proposed that such stimulators could be implemented using optogenetics and be implanted for long term use. More generally, the notion of precise neural control raises questions about the amount of information that can be carried using neurons and the reliability of that information. Information-theoretic analysis of a single ChR receptor in \cite{Eckford2016a} showed that it has a remarkably high capability of receiving information. However, the information propagated by neurons is typically observed via the pulses that fire and not the behavior of individual receptors. There is no one universal method for neurons to encode information, but researchers typically measure the number and timing of fired pulses or ``spikes''; see \cite{Houghton2011}. Examples of the importance of timing include \cite{Srivastava2017}, where neural spike timing patterns in songbirds were manipulated with millisecond-scale variations to control respiratory behavior.

In this paper, we measure how effectively we could use optogenetics to externally stimulate a spike train to match a ``target'' spike train. We model an ideal neuron that is charged by a light source. Inspired by the metrics-based approach reviewed in \cite{Houghton2011} for natural responses to stimuli, we compare the target and generated sequences by measuring the ``distance'' between them. From our perspective, the distance between the sequences is a \emph{distortion} between the generated train and the target train. In practice, there should be a \emph{threshold distortion} below which the pertinent information in the spike train can be recovered. We apply simple filter-based metrics and measure the timing distortion as the root mean square error (RMSE) between the filtered output of the two sequences. We derive the distortion and approximate the expected distortion when matching randomly-generated target sequences.

The rest of the paper is organized as follows. Section~\ref{section_model} describes the neuron firing model. We derive the average sequence distortion in Section~\ref{sec_distortion}. We verify our derivations with simulations in Section~\ref{sec_results} and conclude in Section~\ref{sec_conclusions}.

\section{System Model}
\label{section_model}

We consider a system that has a light source and a single neuron with multiple light-sensitive receptors on its surface. The light source can illuminate the neuron, which opens the receptor ion channels on its surface and increases its internal potential until it fires. For the sake of analysis, we will make several (mostly realistic) assumptions about this process:
\begin{enumerate}
    \item The light source is binary, i.e., either on or off. This assumption can be appropriate for lasers or LEDs, which are both common in optogenetics; see \cite{Fenno2011}.
    \item The process of receptors opening is stochastic (e.g., see \cite{Nagel2003}), but we assume that the current is equal to its expected value $\ion$ when the light is on. This assumption is appropriate if the number of receptors on the neuron is sufficiently large.
    \item The neuron uses the integrate-and-fire model from \cite{Abbott1999} with capacitance $C$ and threshold $\tau$. The integrate-and-fire model provides analytical simplicity at the expense of some fidelity; generalization of this model is possible.
    \item Time is discretized into slots of $\dt$ that are shorter than the time necessary to charge the neuron. Specifically, there exists an integer $\nMin$ such that the integrate-and-fire threshold satisfies
    \begin{equation}
        \label{eqn:ThresholdCondition}
        \tau = \nMin \frac{I_{\mathrm{on}} \Delta t}{C} .
    \end{equation}
    In continuous time, the minimum firing time is $\tMin = \nMin \dt$, but we will focus on the discrete time model.
\end{enumerate}

In the following, we give some interpretive statements about our assumptions. Let $V(t)$ represent the neuron potential as a function of time.
From assumptions 1-3, when the light is on, the neuron behaves as an ideal capacitive circuit with a current $\ion$. Thus, if the light is on from time $t_1$ to $t_2$, then the change in potential $V(t_2) - V(t_1)$ is
\begin{equation}
    \label{eqn:ChangeInPotentialContinuous}
    V(t_2)-V(t_1) = \frac{1}{C} \int_{t_1}^{t_2} I_{\mathrm{on}} \:dt = \frac{I_{\mathrm{on}} (t_2-t_1)}{C} .
\end{equation}
If the light is off, the current is zero, so $V(t_2)-V(t_1) = 0$.

From assumption 3, once $V(t)$ exceeds the threshold $\tau$, the neuron fires and $V(t)$ is immediately reset to zero.

\begin{figure}[!tb]
	\centering
	\includegraphics[width=\linewidth]{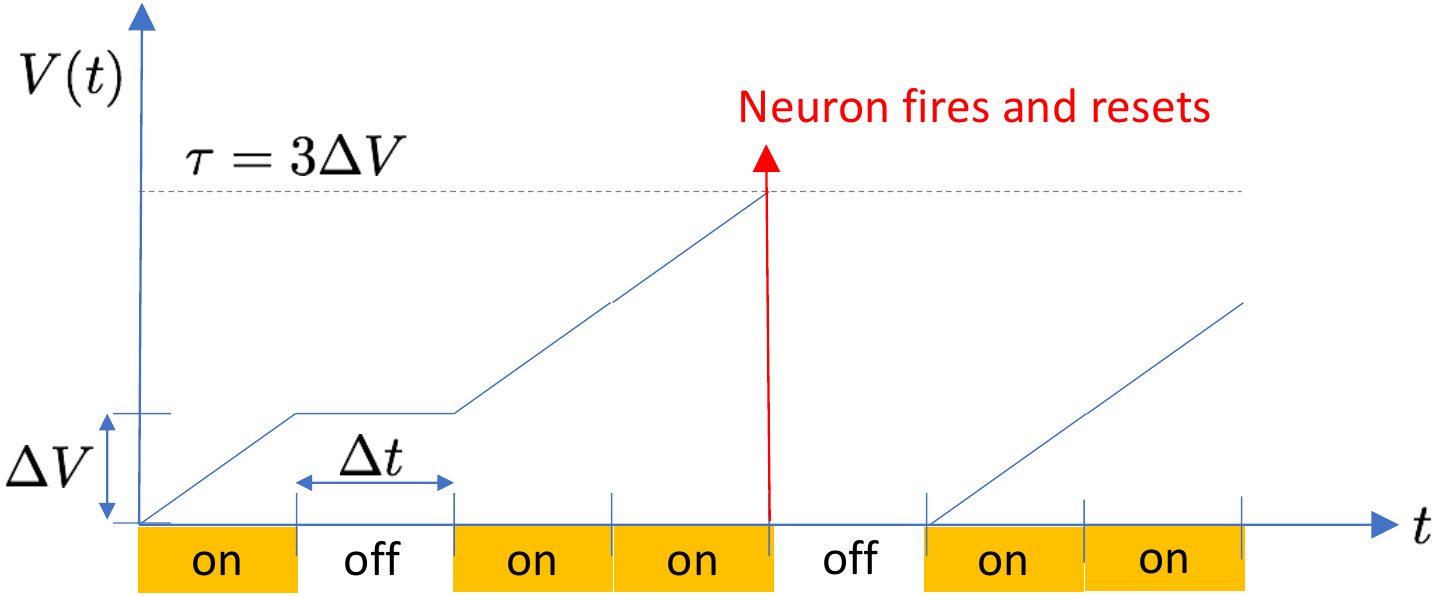}
	\caption{Illustration of the neuron model with integrate-and-fire. In this example, $\nMin = 3$, i.e., spikes must be separated by at least $3 \Delta t$. In each interval when the light source is on, the voltage increases by $\Delta V$. Once the threshold $\tau = 3 \Delta V$ is reached, the neuron fires and resets to $V(t) = 0$. }
	\label{fig:ModelDiagram}
\end{figure}

From assumption 4,  we will say that the light is synchronized with the discrete-time clock, and is either on or off for an entire interval $\Delta t$. Then from (\ref{eqn:ChangeInPotentialContinuous}) we can define $\Delta V$ as
\begin{equation}
    \label{eqn:DeltaV}
    \Delta V = V(t + \Delta t) - V(t) = \frac{I_{\mathrm{on}} \Delta t}{C} 
\end{equation}
when the light source is on.
Finally, since $\tau = \nMin \Delta V$ from (\ref{eqn:ThresholdCondition}), the light must be on for $\nMin$ slots in order for the neuron to fire. This is depicted in Fig. \ref{fig:ModelDiagram}.

\begin{figure}[!tb]
	\centering
	\includegraphics[width=\linewidth]{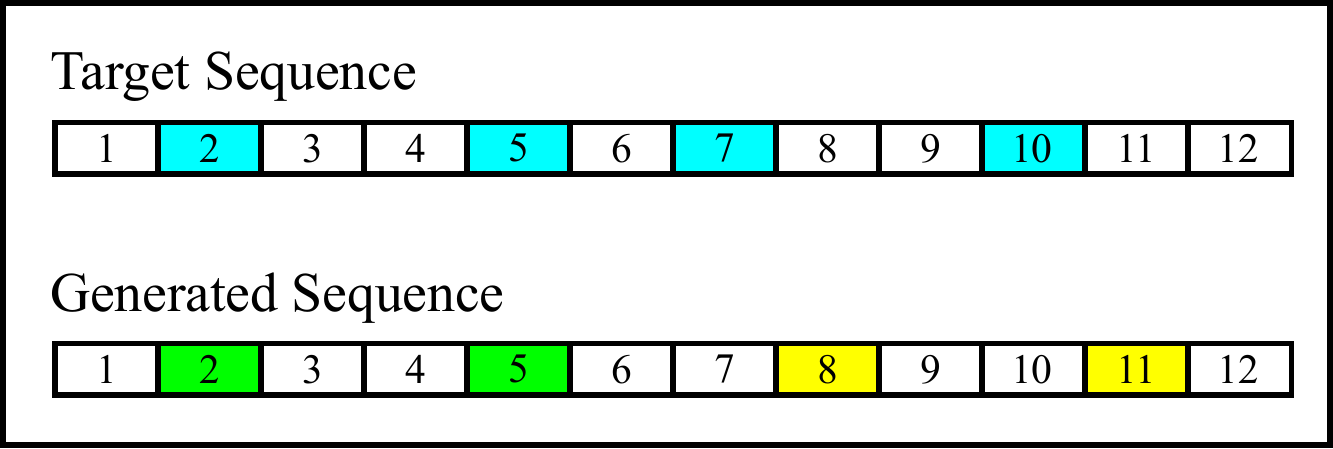}
	\caption{Example of target sequence matching in discrete time, where the generated sequence is constrained by a charging time of $\nMin = 3$~slots. Slots are labeled chronologically and colored when there is a spike at the start of the slot. The target sequence has 4 pulses (shown in blue). The generated sequence can match the first 2 pulses (green), but the final 2 pulses (yellow) each have a delay of one slot.}
	\label{fig_matching_example}
\end{figure}

In this work, we consider using the light source to generate a train of spikes to match a target sequence, where we are constrained by the time it takes to charge and fire the neuron. This matching problem is demonstrated in Fig.~\ref{fig_matching_example}. We suppose that there is a target spike train $\targetSeq$ that we define by the timing of its individual spikes, i.e., $\targetSeq = \{\target{1}, \target{2}, \ldots,\target{\targetLen}\}$, where $\target{i}$ is the time slot when the $i$th spike fires. We assume that $\targetSeq$ is known \emph{a priori}. We may not be able to generate $\targetSeq$ perfectly, e.g., it may be the superposition of spike trains from multiple neurons, but instead we use the light source to generate the sequence $\genSeq = \{\gen{1}, \gen{2}, \ldots,\gen{\genLen}\}$. The only constraint on neuron firing times in $\genSeq$ that we consider is the time that it takes to charge up the neuron to the threshold voltage $\threshold$. Since $\targetSeq$ is known \emph{a priori}, we can turn on the light source and begin charging the neuron \emph{before} the corresponding target firing time. As long as a given target spike is at least $\nMin$ slots since the previous spike, then we can generate a corresponding spike at the precise target time. This is a simplified and ideal generation model but it facilitates tractable analysis. An interesting variation for future work is to refrain from matching a spike if that enables us to match future spikes, e.g., omit generating a spike to match the target in the 7th time slot in Fig.~\ref{fig_matching_example} so that we can match the target in the 10th time slot.

Our goal in the remainder of this paper is to measure the ``distance'' of the sequence $\genSeq$ from the sequence $\targetSeq$, subject to a distance or distortion measure $\distortFcn{\targetSeq,\genSeq}$. Obviously, if all spikes in $\targetSeq$ are separated by at least $\nMin$ slots, then in our model we can generate $\genSeq = \targetSeq$ and we should have $d(\targetSeq,\genSeq) = 0$.

\section{RMSE in Generated Spike Trains}
\label{sec_distortion}

In this section, we present the filter-based metric model from \cite{Houghton2011} for comparing spike trains and apply it to measure the root mean square error (RMSE) of the timing distortion between the target sequence $\targetSeq$ and the generated sequence $\genSeq$. We then simplify the RMSE when the filter has a length of 1 or 2 time slots and derive the expected distortion.

\subsection{Filter-Based Model from \cite{Houghton2011}}

There is no one universal method to encode information in a spike train, but in this work we focus on a filter-based metric that enables some discretion in how to measure the distortion. We begin by mapping the spike trains $\targetSeq$ and $\genSeq$ onto the vector space of functions; see \cite{Houghton2011} for a more general discussion and additional examples. A discrete time model of the $\ell^\mathrm{p}$ norm distortion in \cite[Eq.~(17)]{Houghton2011} between sequences $\targetSeq$ and $\genSeq$ is
\begin{equation}
\label{eqn_distortionDT_lp_norm}
\distortFcnDT{\targetSeq,\genSeq} = \left(\sum_n\left|\mapDT{n;\targetSeq} - \mapDT{n;\genSeq}\right|^p \right)^{1/p},
\end{equation}
where $n$ is the time index and $\mapDT{n;\cdot}$ is the mapping function that maps a sequence to a vector space.  We use a filter function with a kernel $\kernelDT{n}$, such that the sequence $\targetSeq$ maps as
\begin{equation}
\label{eqn_filter_function}
\mapDT{n;\targetSeq} = \sum_{i=1}^\targetLen \kernelDT{n-\target{i}}.
\end{equation}

For ease of analysis, we are interested in finite-length kernels. Other kernels considered in \cite{Houghton2011} include the Gaussian filter and the exponential filter, but they are outside the scope of this work. Notably, the exponential filter associates the mapping function with a neuron's post-synaptic conductance.

\subsection{Filter-Based Metric with RMSE}

We now focus on the $\ell^2$ norm, i.e., the Euclidean distance or RMSE between the two sequences in vector space, which we will see is sensitive to the timing of the individual spikes in $\targetSeq$ and $\genSeq$. From (\ref{eqn_distortionDT_lp_norm}) and (\ref{eqn_filter_function}), the distortion can be written as
\begin{align}
    \distortFcnDT{\targetSeq,\genSeq} = & \left(\sum_n\left(\sum_{i=1}^\targetLen \kernelDT{n-\target{i}} - \sum_{i=1}^\genLen \kernelDT{n-\gen{i}}\right)^2\right)^\frac{1}{2} \nonumber \\
    = & \vast(\sum_n\vast[\left(\sum_{i=1}^\targetLen \kernelDT{n-\target{i}}\right)^2 + \left(\sum_{i=1}^\genLen \kernelDT{n-\gen{i}}\right)^2 \nonumber \\
    & - 2\sum_{i=1}^\targetLen \sum_{j=1}^\genLen \kernelDT{n-\target{i}}\kernelDT{n-\gen{j}} \vast]\vast)^\frac{1}{2} \nonumber \\
    = & \vast(\underbrace{\sum_{i=1}^\targetLen \sum_n\kernelDTSq{n-\target{i}}}_\text{Target energy} + \underbrace{\sum_{i=1}^\genLen\sum_n \kernelDTSq{n-\gen{i}}}_\text{Generated energy} \nonumber \\
    & + \underbrace{2\sum_{i=1}^\targetLen\sum_{j=i+1}^\targetLen\sum_n \kernelDT{n-\target{i}} \kernelDT{n-\target{j}}}_\text{Target density} +  \nonumber \\
    & + \underbrace{2\sum_{i=1}^\genLen\sum_{j=i+1}^\genLen\sum_n \kernelDT{n-\gen{i}} \kernelDT{n-\gen{j}}}_\text{Generated density} \nonumber \\
    & - \underbrace{2\sum_{i=1}^\targetLen \sum_{j=1}^\genLen\sum_n \kernelDT{n-\target{i}}\kernelDT{n-\gen{j}}}_\text{Overlap measure} \vast)^\frac{1}{2},
    \label{eqn_l2_distortion}
\end{align}
where we label the terms in (\ref{eqn_l2_distortion}). The two \emph{energy} terms describe the energy of the two filtered sequences. The \emph{density} terms describe the proximity of the individual spikes in each sequence to the other spikes in the \emph{same} sequence. The \emph{overlap measure} describes the proximity of the individual spikes in $\genSeq$ to the spikes in $\targetSeq$. It can be shown, as expected, that the distortion is minimized to $d(\targetSeq,\genSeq) = 0$ when the overlap measure is maximized, i.e., when $\genSeq = \targetSeq$.

We have not yet placed any constraints on the form of the kernel $\kernelDT{n}$ or the two sequences. To simplify the distortion measure, we now impose that we generate a sequence of the same length as the target sequence, i.e., $\genLen = \targetLen$, such that we can write the timing of each spike in $\genSeq$ as $\gen{i} = \target{i} + \offset{i}$, where $\offset{i}$ is the offset of the $i$th generated spike from the target time. Thus, we can write the distortion as
\begin{align}
\distortFcnDT{\targetSeq,\genSeq} = 
    & \vast(2\sum_{i=1}^\targetLen \sum_n\kernelDTSq{n-\target{i}} \nonumber \\
    & +  2\sum_{i=1}^\targetLen\sum_{j=i+1}^\targetLen\Bigg(\sum_n \kernelDT{n-\target{i}} \kernelDT{n-\target{j}} \nonumber \\
    & + \sum_n \kernelDT{n-\target{i}-\offset{i}} \kernelDT{n-\target{j}-\offset{j}}\Bigg) \nonumber \\
    & - 2\sum_{i=1}^\targetLen \sum_{j=1}^\targetLen\sum_n \kernelDT{n-\target{i}}\kernelDT{n-\target{j}-\offset{j}} \vast)^\frac{1}{2}.
    \label{eqn_l2_distortion_offset}
\end{align}

Our notion of ``proximity'' between spikes when measuring the overlap or density of sequences is particularly sensitive to the length of the kernel $\kernelDT{n}$. To explore this further, we next consider kernels of length $\pulseDT = 1$ and $\pulseDT = 2$ discrete time slots. Furthermore, we assume for simplification that the spike times in the target sequence $\targetSeq$ are all unique (i.e., there is no more than one spike in a given slot), and without loss of generality that they are sorted in increasing order.

\subsubsection{Distortion with Kernel of Length 1}
\label{sec_distortion_kernel1}

If $\pulseDT = 1$, then the density of the target sequence must be 0, i.e., there is no partial overlap between spikes in the same sequence. Every spike in the generated sequence either perfectly matches or misses a target spike; partial overlap between the two sequences is not possible. Relative to other kernel lengths, the \emph{overlap} term is minimized. From the perspective of matching $\targetSeq$ with $\genSeq$, this distortion measure discards every generated spike that does not align with any target spike and the severity of the misalignment does not matter (unlike, for example, a distortion metric that measures the delay). In other words, the timing of the spikes must be \emph{perfectly synchronous} to have $d(\targetSeq,\genSeq) = 0$.

Applying a Kronecker delta kernel, where $\kernelDT{n} = \delta[n]$, the distortion in (\ref{eqn_l2_distortion_offset}) becomes
\begin{equation}
\distortFcnDT{\targetSeq,\genSeq} = \left(2\targetLen - 2\sum_{i=1}^\targetLen \sum_{j=1}^\targetLen \left(\target{i} \stackrel{?}{=} \target{j} + \offset{j}\right) \right)^\frac{1}{2},
\label{eqn_l2_distortion_simplified}
\end{equation}
where $\left(\target{i} \stackrel{?}{=} \target{j} + \offset{j}\right)$ is an indicator function with value 1 if the equality is true. Eq.~(\ref{eqn_l2_distortion_simplified}) is in a form that we can readily evaluate from the target sequence $\targetSeq$ and the offset sequence $\offsetSeq = \{\offset{1},\offset{2},\ldots,\offset{\targetLen}\}$. We note that (\ref{eqn_l2_distortion_simplified}) does \emph{not} impose that a generated spike has to align with its corresponding target in order to prevent distortion due to that spike, nor does it make any assumptions about the individual offset values $\offset{j}$ (i.e., they could be any integer). However, since we cannot generate multiple spikes simultaneously (because we must re-charge the neuron after every spike generation), we know that the indicator function can only be true for at most one value of $j$ for every value of $i$ (and vice versa).

Next, we consider the expected distortion measure $\distortFcnDT{\targetSeq,\genSeq}$. In order to do so, we must make additional assumptions about the target sequence and the offset sequence. We impose that each offset $\offset{j}$ must be a \emph{delay}, such that $\offset{j} \ge 0$. We also assume that the target sequence is sparse, such that it has no more than 2 spikes within any interval of $2\nMin$ slots, where $\nMin$ is the minimum number of slots between spikes in $\genSeq$. Thus, each $\offset{j}$ will only depend on the values of the corresponding $\target{j}$ and $\target{j-1}$, such that we are never waiting to generate more than 1 spike at a time. By imposing these assumptions, a generated spike that occurs at the same time as a target spike must be intended for that target, and we can approximate the distortion in (\ref{eqn_l2_distortion_simplified}) as
\begin{equation}
\distortFcnDT{\targetSeq,\genSeq} \approx \left(2\targetLen - 2\sum_{i=1}^\targetLen \left(\offset{i} \stackrel{?}{=} 0\right) \right)^\frac{1}{2}.
\label{eqn_l2_distortion_simplified_low}
\end{equation}

Let us consider whether the assumptions for (\ref{eqn_l2_distortion_simplified_low}) prevent us from satisfying $\left(\target{i} \stackrel{?}{=} \target{j} + \offset{j}\right)$ in (\ref{eqn_l2_distortion_simplified}) when $\offset{i} \ne 0$. We can prove by contradiction that this is true. If $\offset{i} \ne 0$, then $\left(\target{i} \stackrel{?}{=} \target{j} + \offset{j}\right)$ could only be true for some $i \ne j$. We've imposed that $\offset{i}$ must be non-negative, so we can only consider $j < i$. The most recent case to satisfy the indicator function would be $j = i - 1$, such that $\target{i} = \target{i-1} + \offset{i-1}$. However, if $\offset{i-1} \ne 0$, then $\target{i-1} - \target{i-2} < \nMin$, and the timing of the $(i-2)$th, $(i-1)$th, and $i$th spikes violates our assumption that there can be no more than 2 target spikes within any interval of $2\nMin$ slots. So, we only need to look for cases of $\offset{i} = 0$, i.e., when spikes are generated with no delay, which leads to (\ref{eqn_l2_distortion_simplified_low}). From (\ref{eqn_l2_distortion_simplified_low}), we see that the distortion measure is reduced for each spike that is generated without delay.

We are able to determine the expectation of the approximate distortion in (\ref{eqn_l2_distortion_simplified_low}). First, we need the probability that $\offset{i} = 0$. Since $\targetSeq$ is increasing, the first offset $\offset{1} = 0$. For $i>1$, we know that $\offset{i} = 0$ if there is sufficient separation between the current and previous target spikes, i.e., if $\target{i} - \target{i-1} \ge \nMin$. To maximize the entropy of the target sequence, we assume that the number of slots separating consecutive target spikes follows a geometric distribution with probability $\probTarget$; see \cite{Ross2009}. Thus, we can estimate the probability that $\offset{i} = 0$ as
\begin{align}
\Pr(\offset{i} = 0) \approx & \Pr(\target{i} - \target{i-1} \ge \nMin) \nonumber \\
    = & \left(1-\probTarget\right)^{\nMin-1},
\label{eqn_l2_distortion_cdf}
\end{align}
and so $(\offset{i} \stackrel{?}{=} 0), i > 1$, is a Bernoulli random variable with success probability $\left(1-\probTarget\right)^{\nMin-1}$. Furthermore, the summation $\bDelayRV{}= \sum_{i=2}^\targetLen(\offset{i} \stackrel{?}{=} 0)$, which is the number of target spikes (after the initial spike) that we can generate with no delay, is a Binomial random variable with $\targetLen-1$ trials and value $\bDelayRVval{}$. From the properties of functions of random variables (see \cite{Proakis2000}), we can then write the \emph{expected} distortion as follows:
\begin{align}
\distortFcnDTAvg{\targetSeq,\genSeq} = &\, \mathbf{E}\left[\distortFcnDT{\targetSeq,\genSeq}\right] \approx
\sum_{\bDelayRVval{}=0}^{\targetLen-1} \left(2\targetLen - 2 - 2\bDelayRVval{} \right)^\frac{1}{2} \pmfFcn{\bDelayRVval{}} \nonumber \\
= & \sum_{\bDelayRVval{}=0}^{\targetLen-1} \left(2\targetLen - 2 - 2\bDelayRVval{} \right)^\frac{1}{2}\binom{\targetLen-1}{\bDelayRVval{}} \nonumber \\
&\times \left(1-\probTarget\right)^{\bDelayRVval{}(\nMin-1)} \left(1-\left(1-\probTarget\right)^{\nMin-1}\right)^{\targetLen-1-\bDelayRVval{}},
\label{eqn_l2_distortion_mean}
\end{align}
where $\pmfFcn{\bDelayRVval{}}$ is the probability mass function (PMF) of the random variable $\bDelayRV{}$.

\subsubsection{Distortion with Kernel of Length 2}
\label{sec_distortion_kernel2}

If $\pulseDT > 1$, then partial overlap between spikes is possible, and the degree of distortion is less sensitive to the precise alignment of the spikes. For example, let us consider the case $\pulseDT = 2$, such that $\kernelDT{n} = \kernelVal{0}\delta[n]+\kernelVal{1}\delta[n-1]$. The degree of overlap within a sequence is still limited and can only occur between consecutive spikes, i.e., the $i$th spike can overlap the $(i+1)$th spike but not the $(i+2)$th. By applying this constraint, it can be shown that the distortion in (\ref{eqn_l2_distortion_offset}) simplifies to
\begin{align}
\distortFcnDT{\targetSeq,\genSeq} = & \Bigg(2\targetLen(\kernelVal{0}^2 + \kernelVal{1}^2) +  2\sum_{i=1}^{\targetLen-1} \kernelVal{0}\kernelVal{1}(\target{i+1} \stackrel{?}{=} \target{i} +1) \nonumber \\
    & - 2\sum_{i=1}^\targetLen \sum_{j=1}^\targetLen\Big[(\kernelVal{0}^2 + \kernelVal{1}^2)(\target{i} \stackrel{?}{=} \target{j} +\offset{j}) \nonumber \\
    & + \kernelVal{0}\kernelVal{1}(\target{i} \stackrel{?}{=} \target{j}+\offset{j} \pm 1)\Big] \Bigg)^\frac{1}{2},
    \label{eqn_l2_distortion_simplified_width2}
\end{align}
which is exact. To simplify (\ref{eqn_l2_distortion_simplified_width2}) and approximate the mean distortion, we assume that every offset must be a delay and that the target sequence has no more than 2 spikes within any interval of $2(\nMin+1)$ slots. Then, as in the case where $\pulseDT=1$, the only slots that could have overlap between $\targetSeq$ and $\genSeq$ is when a spike can be generated at the same time as its corresponding target with no delay, i.e., $\offset{i} = 0$. We also assume that the number of slots separating consecutive target spikes follows a geometric distribution with probability $\probTarget$. Applying these assumptions to (\ref{eqn_l2_distortion_simplified_width2}) leads to
\begin{align}
\distortFcnDT{\targetSeq,\genSeq} \approx
    & \Bigg(2\targetLen(\kernelVal{0}^2 + \kernelVal{1}^2) +  2\kernelVal{0}\kernelVal{1}
    \sum_{i=1}^{\targetLen-1} (\target{i+1} \stackrel{?}{=} \target{i} +1) \nonumber \\
    & - 2(\kernelVal{0}^2 + \kernelVal{1}^2)\left(1+\sum_{i=2}^\targetLen(\target{i} - \target{i-1} \stackrel{?}{\ge} \nMin)\right) \Bigg)^\frac{1}{2}.
    \label{eqn_l2_distortion_simplified_low_width2}
\end{align}

The summation $\bDelayRV{1}= \sum_{i=1}^{\targetLen-1}(\target{i+1} \stackrel{?}{=} \target{i} +1)$ is a Binomial random variable with $\targetLen-1$ trials, success probability $\probTarget$, and value $\bDelayRVval{1}$. Analogously to the case where $\pulseDT = 1$, the summation $\bDelayRV{2}= \sum_{i=2}^\targetLen(\target{i} - \target{i-1} \stackrel{?}{\ge} \nMin)$ is a Binomial random variable with $\targetLen-1$ trials, success probability $\left(1-\probTarget\right)^{\nMin-1}$, and value $\bDelayRVval{2}$. Qualitatively, $\bDelayRV{1}$ is the number of target spikes that are in the slot \emph{immediately} following the previous target spike, and $\bDelayRV{2}$ is the number of target spikes that are \emph{at least} $\nMin$ slots after the previous target spike.

Next, we approximate the expected distortion. The difference between this case and $\pulseDT = 1$ is that we must determine the joint PMF $\pmfFcn{\bDelayRVval{1},\bDelayRVval{2}}$ of two \emph{dependent} Binomial random variables $\bDelayRV{1}$ and $\bDelayRV{2}$. We derive the joint PMF using the multiplicative rule for joint probabilities, i.e.,
\begin{equation}
\pmfFcn{\bDelayRVval{1},\bDelayRVval{2}} = \pmfFcn{\bDelayRVval{2}|\bDelayRVval{1}} \pmfFcn{\bDelayRVval{1}},
\end{equation}
where $\pmfFcn{\bDelayRVval{1}}$ is the Binomial PMF with $\targetLen-1$ trials and success probability $\probTarget$. Given knowledge of $\bDelayRVval{1}$, there are fewer trials for the occurrence of $\bDelayRVval{2}$ (reduced to $\targetLen-1-\bDelayRVval{1}$) but a higher success probability $(1-\probTarget)^{\nMin-2}$. Thus, we can write the expected distortion as
\begin{align}
\distortFcnDTAvg{\targetSeq,\genSeq} \approx &
\sum_{\bDelayRVval{1}=0}^{\targetLen-1} \sum_{\bDelayRVval{2}=0}^{\targetLen-1-\bDelayRVval{1}} \big[(2\targetLen - 2 - 2\bDelayRVval{2})(\kernelVal{0}^2 + \kernelVal{1}^2) \nonumber \\
& + 2\kernelVal{0}\kernelVal{1}\bDelayRVval{1}\big]^\frac{1}{2} \pmfFcn{\bDelayRVval{1},\bDelayRVval{2}} \nonumber \\
= & \sum_{\bDelayRVval{1}=0}^{\targetLen-1} \binom{\targetLen-1}{\bDelayRVval{1}}\probTarget^{\bDelayRVval{1}}(1-\probTarget)^{\targetLen-1-\bDelayRVval{1}} \nonumber \\
& \times \!\!
\sum_{\bDelayRVval{2}=0}^{\targetLen-1-\bDelayRVval{1}}\!\!\! \left[(2\targetLen - 2 - 2\bDelayRVval{2})(\kernelVal{0}^2 + \kernelVal{1}^2) + 2\kernelVal{0}\kernelVal{1}\bDelayRVval{1}\right]^\frac{1}{2} \nonumber \\
& \times \binom{\targetLen-1-\bDelayRVval{1}}{\bDelayRVval{2}}(1-\probTarget)^{\bDelayRVval{2}(\nMin-2)} \nonumber \\
& \times\left(1-(1-\probTarget)^{\nMin-2}\right)^{\targetLen-1-\bDelayRVval{1}-\bDelayRVval{2}},
\label{eqn_l2_distortion_mean_width2}
\end{align}
which is exact for a low target spike density. However, its verbosity makes it less intuitive than (\ref{eqn_l2_distortion_mean}), i.e., when $\pulseDT=1$.

\subsubsection{Consideration of Longer Kernels}

A comparison between (\ref{eqn_l2_distortion_simplified}) and (\ref{eqn_l2_distortion_simplified_width2}) demonstrates the increase in complexity when we need to account for partial overlap between filtered spikes in the same or different sequences. However, the approximations that we applied to simplify the distortion and derive the expected distortion suggest that a tractable expression to approximate the distortion for any arbitrary filter length is feasible. We leave such consideration for future work.

\section{Numerical Results}
\label{sec_results}

In this section, we evaluate the performance of the RMSE timing distortion metric that we derived in Section~\ref{sec_distortion}. We assume throughout this section that the target sequence $\targetSeq$ has $\targetLen = 10$ spikes that are generated with success probability $\probTarget \in [10^{-3},1]$ in every discrete time slot. We generate target sequences by simulating the geometric distribution, and all simulation results are generated by averaging over at least $10^4$ sequences for each $\probTarget$, such that every plotted simulation point has an insignificant confidence interval. We consider a minimum charging time $\tMin=2\,\mathrm{ms}$, which is consistent with the typical neuron recovery time; see \cite[Ch.~45]{Sadava2014}. We use discrete time slots of length $\dt=0.5\,\mathrm{ms}$, so there is a minimum of $\nMin=4$ slots between generated spikes.

For the filter of length $\pulseDT = 1$, we consider the Kronecker delta kernel so that we can apply the results from Section~\ref{sec_distortion_kernel1}. For the filter of length $\pulseDT = 2$, we consider filter coefficients $\{\kernelVal{0},\kernelVal{1}\} = \{\sqrt{0.5},\sqrt{0.5}\}$, so that each coefficient is weighted equally and the sum of the squares of the coefficients is equal to that of the Kronecker delta.

\begin{figure}[!tb]
	\centering
	\includegraphics[width=\linewidth]{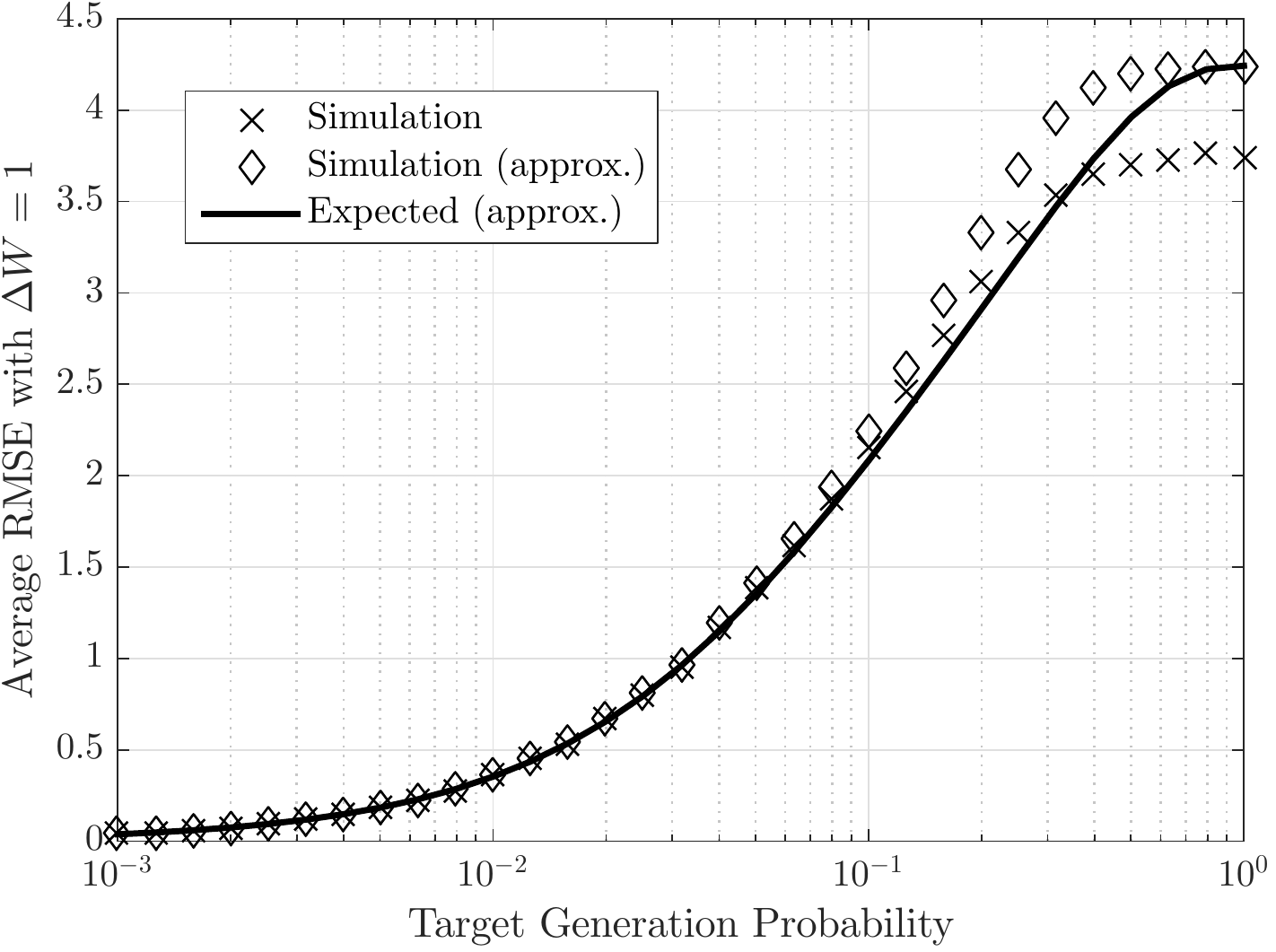}
	\caption{Average RMSE $\distortFcnDTAvg{\targetSeq,\genSeq}$ as a function of the target firing probability $\probTarget$ in each time slot. The expected analytical curve is evaluated from (\ref{eqn_l2_distortion_mean}). The actual observed distortion is calculated using (\ref{eqn_l2_distortion_simplified}), and the approximate observed distortion is calculated using (\ref{eqn_l2_distortion_simplified_low}).}
	\label{fig_distortion_mse_length1}
\end{figure}

In Fig.~\ref{fig_distortion_mse_length1}, we measure the average RMSE as a function of the target spike generation probability $\probTarget$ for the filter of length $\pulseDT = 1$. We use two methods to calculate the simulated distortion. The true simulated distortion is measured using (\ref{eqn_l2_distortion_simplified}). We approximate the simulated distortion using (\ref{eqn_l2_distortion_simplified_low}), where we assume that $\probTarget$ is sufficiently low, i.e., that $\targetSeq$ is sparse. The expected analytical curve is plotted using (\ref{eqn_l2_distortion_mean}) and also assumes that $\targetSeq$ is sparse.

\begin{figure}[!tb]
	\centering
	\includegraphics[width=\linewidth]{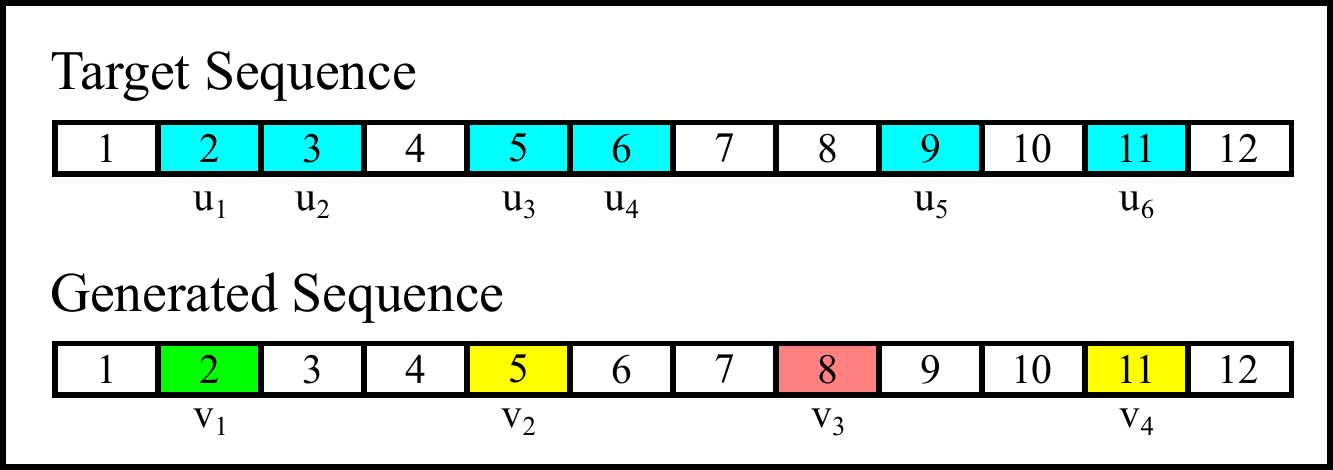}
	\caption{Example of target sequence matching where the target sequence $\targetSeq$ has a high spike generation probability and the generated sequence $\genSeq$ is constrained by a charging time of $\nMin=3$~slots. $\targetSeq$ has 6 pulses (at start of slots shown in blue). $\genSeq$ can match the first pulse (green), but the spikes in the $5$th and $11$th slots match \emph{future} spikes that are not the corresponding target spikes (yellow). The spike in the $8$th slot does not match any spike (red). The spikes in $\genSeq$ to match $\target{5}$ and $\target{6}$ occur after the $12$th slot and are not shown.}
	\label{fig_matching_example_high_density}
\end{figure}

We observe in Fig.~\ref{fig_distortion_mse_length1} that all three curves agree well when $\probTarget < 4\,\%$, such that the timing of a given generated spike in $\genSeq$ primarily depends on the timing of only one previous spike in $\targetSeq$. For $\probTarget \ge 4\,\%$, we often have two or more spikes within an interval of $2\nMin$ slots, i.e., spikes occur sufficiently often that multiple previous spikes in $\targetSeq$ affect the timing of spikes in $\genSeq$. This generally leads to the expected distortion acting as a lower bound. However, for \emph{very} high spike generation probabilities, i.e., $\probTarget \ge 40\,\%$, the true distortion becomes \emph{lower} than predicted by the expected curve. This is because there are so many target spikes in $\targetSeq$ that delayed spikes in $\genSeq$ are likely to occur at the same time as future spikes in $\targetSeq$, i.e., $\target{i} = \target{j} + \offset{j}$ for some $j<i$. Examples of this are shown in Fig.~\ref{fig_matching_example_high_density}. Such occurrences are not accounted for in the derivations of the approximations, where spikes in $\genSeq$ must align with the \emph{corresponding} spikes in $\targetSeq$, but from (\ref{eqn_l2_distortion_simplified}) these asynchronous overlaps lead to a smaller distortion. Thus, the expected distortion is a lower bound in the ``low density'' regime but an upper bound in the ``high density'' regime. We also note that the approximate simulated distortion converges to the expected distortion as $\probTarget \to 1$. This is because \emph{every} spike generated after the initial one has \emph{no overlap} with its corresponding target spike, so both distortion measures are maximized to the same value, i.e., $\distortFcnDTAvg{\targetSeq,\genSeq} = \sqrt{2\targetLen-2} = \sqrt{18}$.

\begin{figure}[!tb]
	\centering
	\includegraphics[width=\linewidth]{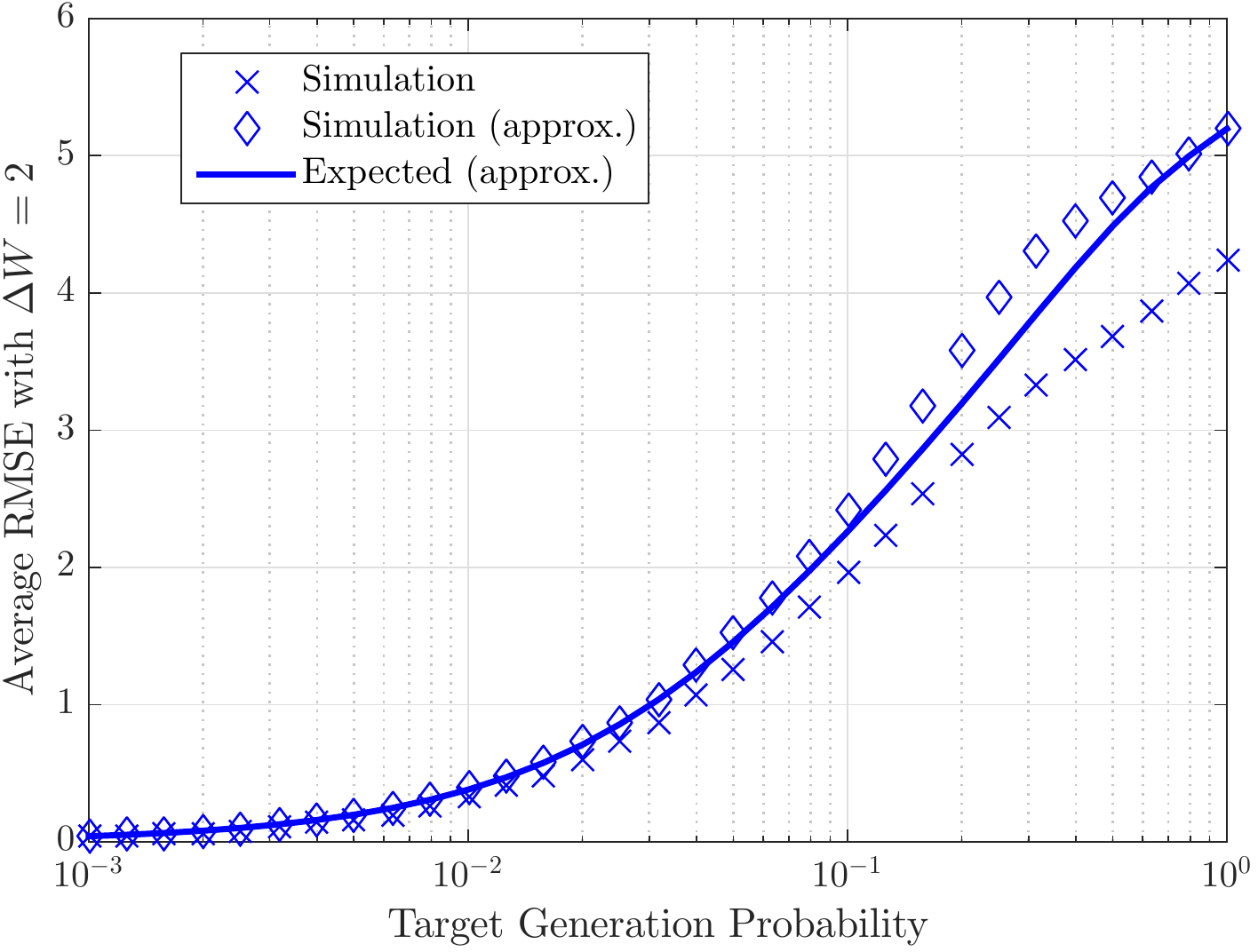}
	\caption{Average RMSE $\distortFcnDTAvg{\targetSeq,\genSeq}$ as a function of the target firing probability  $\probTarget$ in each time slot. The metric kernel has length $\pulseDT=2$ and coefficients $\{\kernelVal{0},\kernelVal{1}\} = \{\sqrt{0.5},\sqrt{0.5}\}$. The expected analytical curve is evaluated from (\ref{eqn_l2_distortion_mean_width2}). The actual observed distortion is calculated using (\ref{eqn_l2_distortion_simplified_width2}), and the approximate observed distortion is calculated using (\ref{eqn_l2_distortion_simplified_low_width2}).}
	\label{fig_distortion_mse_length2}
\end{figure}

In Fig.~\ref{fig_distortion_mse_length2}, we measure the average RMSE as a function of the target spike generation probability $\probTarget$ for the filter of length $\pulseDT = 2$. Analogously to the filter of length 1, we use two methods to calculate the simulated distortion. The true simulated distortion is measured using (\ref{eqn_l2_distortion_simplified_width2}). We approximate the simulated distortion using (\ref{eqn_l2_distortion_simplified_low_width2}), where we assume that $\probTarget$ is sufficiently low. The expected analytical curve is plotted using (\ref{eqn_l2_distortion_mean_width2}) and also assumes that $\probTarget$ is sufficiently low.

As in Fig.~\ref{fig_distortion_mse_length1} when $\pulseDT = 1$, Fig.~\ref{fig_distortion_mse_length2} shows that the expected distortion is a lower bound on the approximate simulated distortion. This bound is accurate for low $\probTarget$ (here when $\probTarget < 5\,\%$) and then converges when $\probTarget \to 1$. However, unlike the $\pulseDT = 1$ case, Fig.~\ref{fig_distortion_mse_length2} demonstrates that the expected distortion is an upper bound on the true distortion for \emph{all} $\probTarget$. This is a side effect of the longer filter; non-zero overlap occurs between a target spike in $\targetSeq$ and the corresponding spike generated in $\genSeq$ when the latter is generated with a one-slot delay. Such ``imperfect'' overlap reduces the measure of distortion calculated using (\ref{eqn_l2_distortion_simplified_width2}) but is not accounted for in either the approximate or expected distortion.

\section{Conclusions}
\label{sec_conclusions}

In this paper, we used a simple optogenetic model to externally stimulate a neuron and generate a spike train. We constrained the neuron's charging time and measured the distortion in the spike train as the filtered train's RMSE from a filtered target sequence. We showed that the expected distortion can be accurately predicted when the spike generation rate in the target sequence is sufficiently low.

Ultimately, this is a preliminary work to understand the information that can be carried in a sequence of neuron pulses. Future work includes a more complete statistical description of the distortion and consideration of other filters and distortion metrics. We are also interested in comparing our analysis with experimental neuron firing data, and considering the propagation of stimulated pulses between connected neurons.

\bibliography{NoelEckfordDistortion}

\end{document}